\documentclass[english,a4paper,11pt]{article}
\usepackage[margin=3cm]{geometry}
\usepackage[latin1,english]{inputenc}
\usepackage{babel}
\usepackage{bm}
\usepackage{amsmath,amsthm}
\usepackage{latexsym}
\usepackage{booktabs}
\usepackage[final]{graphicx}
\DeclareGraphicsExtensions{.jpg,.jpeg,.pdf,.png,.mps}
\usepackage{epsfig}
\usepackage[round]{natbib}
\setcitestyle{numbers}
\setcitestyle{square}

\usepackage{rotating}
\usepackage{color}
\usepackage{xcolor}
\usepackage{colortbl}
\usepackage[bitstream-charter]{mathdesign}
\usepackage[T1]{fontenc}
\usepackage{threeparttable}
\usepackage{lipsum}
\usepackage{url} 
\usepackage{hyperref}
\usepackage{lmodern}
\date{}
\hypersetup{hidelinks}

\title{A comparative analysis of machine learning algorithms for predicting probabilities of default}

\author{$\mathrm{Adrian \ Iulian \ Cristescu}^\mathrm{1},  
	\  \mathrm{Matteo \ Giordano}^\mathrm{2}$\\  
	$^\mathrm{1}$\small{\emph{NTT Data Italia}}\\
        $^\mathrm{2}$\small{\emph{ESOMAS Department, University of Turin}}
}

\begin{document}
	\maketitle

\begin{abstract}
Predicting the probability of default (PD) of prospective loans is a critical objective for financial institutions. In recent years, machine learning (ML) algorithms have achieved remarkable success across a wide variety of prediction tasks; yet, they remain relatively underutilised in credit risk analysis. This paper highlights the opportunities that ML algorithms offer to this field by comparing the performance of five predictive models--Random Forests, Decision Trees, XGBoost, Gradient Boosting and AdaBoost--to the predominantly used logistic regression, over a benchmark dataset from Scheule et al. (\textit{Credit Risk Analytics: The R Companion}). Our findings underscore the strengths and weaknesses of each method, providing valuable insights into the most effective ML algorithms for PD prediction in the context of loan portfolios.
\noindent 
\hspace{1cm}
\end{abstract}

%
%
%
%
%

%%%%%%%%%%%%%%%%%%%%%%%%%%%%%%%%%%%%%%%%%%%%%%%
\section{Introduction}\label{Sec:Intro}

The prediction of the probability of default (PD) is a crucial aspect of credit risk management for financial institutions. Credit risk refers to the losses due to the possibility that a borrower will fail to meet their debt obligations, and accurately predicting the PD allows lenders to make informed decisions regarding loan approvals, pricing, and capital reserves. Various covariates in the form of prospective borrower characteristics are known to play different roles in influencing the PD \cite{scheule2017credit}. Historically, logistic regression has been the predominant model for estimating PDs \cite{hosmer2013applied}, offering both ease of use and interpretability. However, this method is known to have limitations in capturing complex relationships between variables.

	In recent years, machine learning (ML) algorithms have achieved remarkable success, demonstrating universal capabilities to model nonlinear interactions among a high number of features. As a result, these methods are gaining traction in credit risk management \cite{noriega2023machine} and more broadly, in finance \cite{weber2024applications}, where they consistently outperform more traditional techniques \cite{shi2022machine}. However, ML algorithms are still relatively underutilised in the field, hindered by a number of challenges that include model transparency and interpretability, reproducibility of results, lack of systematic comparison across methods, and data quality.

	The goal of this paper is to highlight the potential benefits of a more widespread adoption of ML algorithms for PD prediction, addressing some of the aforementioned challenges. We systematically study the performance of five major predictive models, namely Random Forests, Decision Trees, XGBoost, Gradient Boosting and AdaBoost, and compare it to the more traditionally used logistic regression. While accuracy (i.e.~the ratio of correct predictions) is generally the primary evaluation metric, it is increasingly recognised as alone being inadequate for credit risk analysis, where datasets tend to be highly skewed towards non-defaults, and where the cost to the institution of failing to predict a default is vastly greater than the one of a wrongly predicted default. For a more informative evaluation, we base our analysis on the more nuanced metrics suggested by the literature \cite{noriega2023machine}, including recall, precision, F1-score, and the area-under-the-curve (AUC). In our results, the ensemble-type methods (Random Forests, XGBoost, and Gradient Boosting) consistently outperform the simpler models (Decision Trees and logistic regression) across the board, clearly indicating the superior ability of the former to handle complex patterns and imbalanced data.

	Concerning the key interpretability challenge of ML algorithms, recent advancements allow to incorporate a comparative analysis for the contribution of the individual variables, through feature importance metrics, partial dependence plots, and SHAP (SHapley Additive exPlanations) values \cite{gareth2013introduction}. These may furnish to credit risk analysts a framework to better understand how specific features influence the predicted PD, a critical factor when deploying predictive algorithms in highly regulated environments like finance. In our results, certain features are found to consistently play a larger role in predicting defaults, including the number of delinquent credit lines, major derogatory reports, recent credit inquiries, and the debt-to-income ratio. These findings are aligned with the existing research on the determinants of credit risk \cite{qi2023factors}.

	For our comparative study, we used a dataset sourced from Scheule et al.~\cite{scheule2017credit}, which is a well-established and reliable benchmark in the field of credit risk management. The (Python) code for fully reproducing the analysis is available within the first author's M.Sc.~thesis \cite{crisescu24comparative}.

%
%
%
%
%

%%%%%%%%%%%%%%%%%%%%%%%%%%%%%%%%%%%%%%%%%%%%%%%
\section{Data}\label{Sec:Data}

\paragraph{Dataset overview.} We employed a dataset accompanying the monograph by Scheule et al.~\cite{scheule2017credit}, which is a standard benchmark in credit risk analysis. The dataset comprises 5,960 total entries, with 12 covariates summarising key borrower characteristics.

	In the dataset, 3,364 rows are complete and suitable to the analysis following data cleaning procedures. There is however a significant class imbalance, with only 300 observations classified as defaulters and 3,064 as non-defaulters.

\paragraph{Addressing class imbalance.} To ensure a reliable analysis, we incorporated two commonly used preprocessing steps for imbalanced datasets. These form the basis for a robust model training and evaluation.
\begin{enumerate}
\item \textbf{Stratified Dataset Splitting}: In dividing the dataset, we preserved the relative proportions between defaulters and non-defaulters. This ensures that the training, validation, and test sets are representative of each class, maintaining the original class distribution.

\item \textbf{SMOTE (Synthetic Minority Oversampling Technique, \cite{chawla2002smote})}: A technique that generates synthetic samples for the minority class (here, defaulters) to achieve a more balanced distribution, shown to improve the robustness and predictive performances.
\end{enumerate}

%
%
%
%
%

%%%%%%%%%%%%%%%%%%%%%%%%%%%%%%%%%%%%%%%%%%%%%%%
\section{Methods}\label{Sec:Methods}

\paragraph{ML algorithms.} In our analysis, we employed five established methods for supervised learning: Random Forests, Decision Trees, Extreme Gradient Boosting (XGBoost), Gradient Boosting and Adaptive Boosting (AdaBoost); see \cite{bishop2006pattern}. Additionally,  we also considered logistic regression, which represents an interpretable and simple baseline that is still the standard model in credit risk analysis.

\paragraph{Evaluation Metrics.} We systematically compared the performance of the aforementioned models based on four key evaluation metrics for credit risk management \cite{noriega2023machine}: 
\begin{itemize}
\item \textbf{Recall}: Also known as sensitivity, it is defined as the ratio of correctly predicted positive instances to all actual positive instances. It is a measure of the ability of a model to identify true positives (here, defaulters).

\item \textbf{Precision}: It is the ratio of true positive predictions to the total number of positive predictions. It measures the accuracy of the positive class predictions.

\item \textbf{F1-Score}: It is given by the harmonic mean of recall and precision, balancing between the two metrics. This is particularly valuable in the presence of class imbalance.

\item \textbf{AUC (area-under-the-curve)}: It measures the area under the receiver operating characteristic (ROC) curve of a binary classifier, providing an overall summary of its performance in distinguishing between classes.
\end{itemize}

These metrics provide complementary perspectives on model performance, allowing for a nuanced analysis of the strengths and weaknesses of each method. Jointly, they assess the models' ability to correctly identify defaulters and non-defaulters, while balancing the trade-off between false positives and false negatives. Several authors \cite{noriega2023machine} have emphasised the importance of including a more comprehensive evaluation, taking into account the various aspects of the predictive problem in credit risk management, which is characterised by an inherent class imbalance and a severe cost in failing to predict a default.

%
%
%
%
%

%%%%%%%%%%%%%%%%%%%%%%%%%%%%%%%%%%%%%%%%%%%%%%%
\section{Results}\label{Sec:Results}

\paragraph{Performance metrics.} The main results of our analysis are summarised in Table \ref{Tab:Results} below, where the recall, precision, F1-score and AUC values achieved by each method are reported. For each metric, the highest score achieved across the methods is highlighted. We also report the 95\% confidence intervals for recall, precision and F1-scores.

\begin{table}[!ht]
\caption{Performance metrics (and 95\% confidence intervals) for the predictive results}
\label{Tab:Results}
\centering 
		\begin{tabular}{lcccc}
		\toprule
              & Recall & Precision & F1-Score & AUC   \\
            \midrule
              Random Forest &  0.77 (0.72-0.81)  & 0.96 (0.94-0.98) & 0.85 (0.81-0.89) & 
              \underline{0.98}  \\
              Decision Tree &  0.63 (0.58-0.68) & 0.63 (0.58-0.68) & 0.63 (0.58-0.68) & 0.80   \\
              XGBoost & 0.77 (0.72-0.81) & \underline{1 (1-1)}  & \underline{0.87 (0.83-0.90)} 
              & 0.94 \\
              Gradient Boosting & \underline{0.80 (0.76-0.84)} & 0.80 (0.76-0.84) &  0.80 (0.76-0.84) 
                & 0.92   \\
              AdaBoost & \underline{0.80 (0.76-0.84)} & 0.62 (0.56-0.67) & 0.70 (0.65- 0.75)
              & 0.89\\
              Logistic regression & 0.63 (0.58-0.69) & 0.22 (0.17- 0.26) & 0.32	(0.27- 0.37) & 0.7679 \\
		\bottomrule
		\end{tabular}
%  \begin{tablenotes}
%\item[]{\footnotesize \textit{Notes}. Write note here.}
%\end{tablenotes}
\end{table}

\begin{table}[!ht]
\caption{Feature importance ranking across the models}
\label{Tab:Features}
\centering 
		\begin{tabular}{lccccc}
		\toprule
             Top 3 Features & DELINQ & DEROG & NINQ & DEBTINC & VALUE \\
            \midrule
              Random Forest &  1st  & - & - & 3rd & 2nd\\
              Decision Tree &  1st & - & - & 2nd & 3nd \\
              XGBoost & 1st & 2nd & 3rd & - & -\\
              Gradient Boosting & 1st & 2nd & 3rd & -  & - \\
              Logistic Regression & 1st & 2nd & - & 3rd & - \\
		\bottomrule
		\end{tabular}
%  \begin{tablenotes}
%\item[]{\footnotesize \textit{Notes}. Write note here.}
%\end{tablenotes}
\end{table}

	The results display excellent performances by three ensemble methods, Random Forests, XGBoost, and Gradient Boosting. Overall, these achieved comparable values across the board, and consistently outperformed the simpler techniques, Decision Trees and Logistic Regression, as well as AdaBoost. The latter achieved the highest recall (80\%), but its precision was notably lower (62\%), leading to an F1-Score of 70\%. This suggests a tendency to misclassify non-defaulters and a lack of confidence in the prediction of defaulters.

	In summary, our results furnish a strong case for the use of ensemble ML algorithms, in particular Random Forests and boosting techniques, in the context of PD prediction. Compared to more traditional methods, jointly deploying these ML algorithms can furnish superior and nuanced indications for decision making in credit risk analysis.

\paragraph{Feature Importance.} A second key aspect of this study is to investigate the most impactful features used by each method to predict the PD. For the ML algorithms (except for AdaBoost), this was achieved through feature importance scores. Adaboost's interpretability is generally known to be more challenging, as its boosting mechanism, where models are trained sequentially, makes it harder to understand which individual covariates have the largest impact across all iterations. For this reasons, the feature importance analysis was not carried out in this case. For logistic regression, the most impactful covariates were identified based on the size of their respective coefficient. Overall, the top features influencing predictions were found to be (see \cite{scheule2017credit} for a detailed overview on the dataset):
DELINQ, the number of delinquent credit lines; DEROG, the number of major derogatory reports; NINQ, the number of recent credit inquiries; DEBTINC, the debt-to-income ratio; and VALUE, the value of the current property.

While the individual rankings present slight variations, the consistent presence of these variables highlights their critical role in determining default risk. These findings are in substantial agreement with previous studies on the determinants of credit risk \cite{qi2023factors}. Table \ref{Tab:Features} below  details the specific feature rankings across the methods. It can be seen that DELINQ is the most important feature for all the 5 considered models, while DEROG is the second most important for 3 models.

%
%
%
%
%

%%%%%%%%%%%%%%%%%%%%%%%%%%%%%%%%%%%%%%%%%%%%%%%
\section{Conclusions}

This research underscores the potential of ML in contributing to transform credit risk management, offering institutions  a more data-driven and accurate approach to assessing PDs. Our findings demonstrate that ensemble methods outperform simpler ones in terms of both recall and overall discrimination ability. These methods proved effective in identifying defaulters while maintaining high confidence in their predictions, making them suitable for real-world applications in credit risk management. Our work also highlights the importance of robust data preprocessing, through techniques such as stratification and SMOTE to address class imbalance, as well as the interpretability power offered by feature importance analysis.

\paragraph{Practical implications.} The results of our feature importance analysis (Table \ref{Tab:Features}) suggests, in line with the literature, that financial institutions should prioritise the identified features in their decision-making processes. For example, the number of derogatory reports, of delinquent credit lines and of recent credit inquiries could be used to set a flag for high-risk borrowers. Since these variables can be tracked in time, this could offer the opportunity for early detection of risks and useful instruments for restructuring loans during their duration. Finally, these variables should be prioritised in defining individual credit scores.

\paragraph{Remaining challenges.}

	Several areas of improvements remain, both methodological and concerning the specific application of ML techniques in credit risk analysis. In our study, we addressed the class imbalance between defaulters and non-defaulters via SMOTE (cf. Section \ref{Sec:Data}); however, this methods, which is based on interpolating between existing minority class samples, is known to lead to possible generalisation issues. Other alternatives, such as cost-sensitive learning (using weighted loss functions and class-weighted decision trees) could be explored and compared to the results obtained in the present work.

	The interpretability of ML methods is also a fundamental area of active research. In this work, we used importance feature analysis to identify the most important covariates for the PD prediction. Hybrid models, that combine for example a simpler simple logistic regression with more complex methods, could be used for enhanced interpretability.

	Finally, we mention the importance of having access to more recent, reliable and diversified data; and also the opportunity of involving interdisciplinary expertise for feature engineering. These are essential steps to enhance the predictive power and interpretability of the models, realising their full potential in credit risk analysis and  adapting them to the ever-evolving financial landscape.

\paragraph{Acknowledgements.} We are grateful to the Associate Editor and to an anonymous Referee for helpful suggestions that lead to an improvement of this work. M.~G.~has been partially supported by MUR, PRIN project 2022CLTYP4.

\bibliographystyle{unsrt}
\bibliography{References.bib}

%\listofchanges

\end{document}